\documentclass[useAMS,usenatbib]{mn2e}

\usepackage{graphicx}

\title[The implications of particle rotation on the effect of photophoresis]{The implications of particle rotation on the effect of photophoresis}

\author[J. van Eymeren and G. Wurm]{Janine van Eymeren$^{1}$\thanks{E-mail:
janine.vaneymeren@uni-due.de} and Gerhard Wurm$^{1}$\\
$^{1}$Fakult\"at f\"ur Physik, Universit\"at Duisburg-Essen, Lotharstr. 1, 47048 Duisburg, Germany}

\begin{document}

\date{Accepted 2011 October 17.  Received 2011 October 11; in original form 2011 June 16}

\pagerange{\pageref{firstpage}--\pageref{lastpage}} \pubyear{2002}

\maketitle

\label{firstpage}

\begin{abstract}
In a laboratory experiment, water-ice aggregates are trapped in a vacuum chamber at a pressure of 2\,mbar due to photophoresis and thermophoresis. The particles are located between a Peltier element at the bottom at 250\,K and a reservoir of liquid nitrogen at the top at 77\,K. Particle sizes vary between 20\,$\mu \rm m$ and a few hundred $\mu \rm m$. It is found that 95\% of all the particles rotate about their vertical axis. A qualitative model is developed which explains why particles should mainly align to and rotate around the vertical. The results imply that rotation does not decrease the vertical strength of photophoretically driven motion in, e.g., protoplanetary discs.
\end{abstract}


\section{Introduction}
To verify and detail scenarios of protoplanetary disc evolution, of material transport and to interpret astronomical observations, it is crucial to understand the interaction of small dust and ice particles with their surroundings, i.e., gas, light, and other ice and dust particles. Therefore, it would be favourable if free particles were observed and manipulated in laboratory experiments. Fortunately, a technique to do so was recently developed by \citet{Kelling2011}. In their work, they show that micron-sized water-ice particles levitate under the Earth's gravity, trapped in stable positions by photophoresis and thermophoresis. Both phoretic forces are induced by a temperature gradient. In the first case, the temperature gradient exists across the particle (internal gradient), while the gaseous environment is at a constant temperature, in the latter case, the temperature gradient exists in the gaseous environment (external gradient). In the experiments by \citet{Kelling2011}, photophoretic lift is induced by the thermal radiation of a surface at 250\,K.

As already discussed by \cite{Krauss2007}, levitating particles (in their case dust particles) tend to rotate. The same has been observed for the levitating ice particles in the experiment by \cite{Kelling2011}. Therefore, we here use the above mentioned trapping technique to perform a detailed study of the rotation of levitating ice particles which are under the influence of photophoresis, thermophoresis, and gravity. Rotation can have profound implications for the applicability of photophoresis in protoplanetary discs. The analysis of the rotation is important in this context as it has been and still is frequently argued to be an obstacle to photophoresis. The argument is that photophoresis needs an internal temperature gradient which might be levelled out for a rapidly rotating particle causing the photophoretic strength to decrease. Therefore, the main aim of this paper is to observe particle rotation and to estimate the implications on the effect of photophoresis.
\section[]{Experimental setup and first observations}
\begin{figure}
 \centering
 \includegraphics[width=.27\textwidth, bb= 0 0 415 574, clip=]{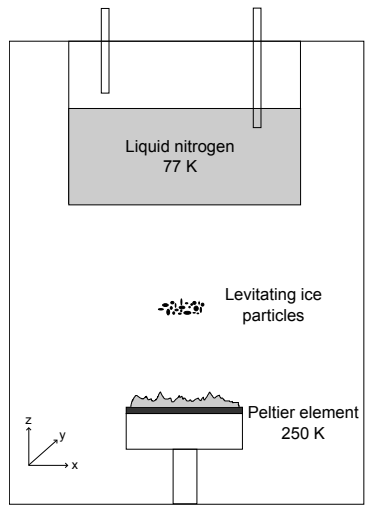}
 \caption{Experimental setup: ice particles are trapped between a Peltier element at 250\,K and a reservoir of liquid nitrogen at 77\,K due to photophoresis induced by the thermal radiation of the Peltier element. The chamber is evacuated to roughly 2\,mbar.}
 \label{Figexpsetup}
\end{figure}

Ice particles are dropped onto a Peltier element cooled down to 250\,K which is situated in a vacuum chamber at room temperature (see Fig.~\ref{Figexpsetup}). A reservoir of liquid nitrogen at 77\,K is placed above the Peltier element at a height of 28\,mm. The chamber is evacuated to roughly 2\,mbar. As soon as the pressure decreases to below 10\,mbar, single ice aggregates begin to rise from the surface of the snow layer. Their different trajectories have been described in detail by \citet{Kelling2011}. We here concentrate on the ice particles which levitate halfway between the Peltier element and the reservoir of liquid nitrogen. We regularly observe the formation of groups of up to a few dozens of particles which are typically located above the edges of the Peltier element. While they are firmly trapped at a fixed height, they still rotate. We do not consider the crystal or aggregate structure of the ice particles. Instead, we simply classify the particles as water-ice. 

The particles vary in size between 20\,$\mu$m up to a few hundred $\mu$m. This range is currently biased by the limited resolution of the optical system and also depends on the particle sizes ejected from the snow placed on the Peltier element.

\begin{figure}
 \centering
 \includegraphics[width=.48\textwidth, bb= 14 15 507 174, clip=]{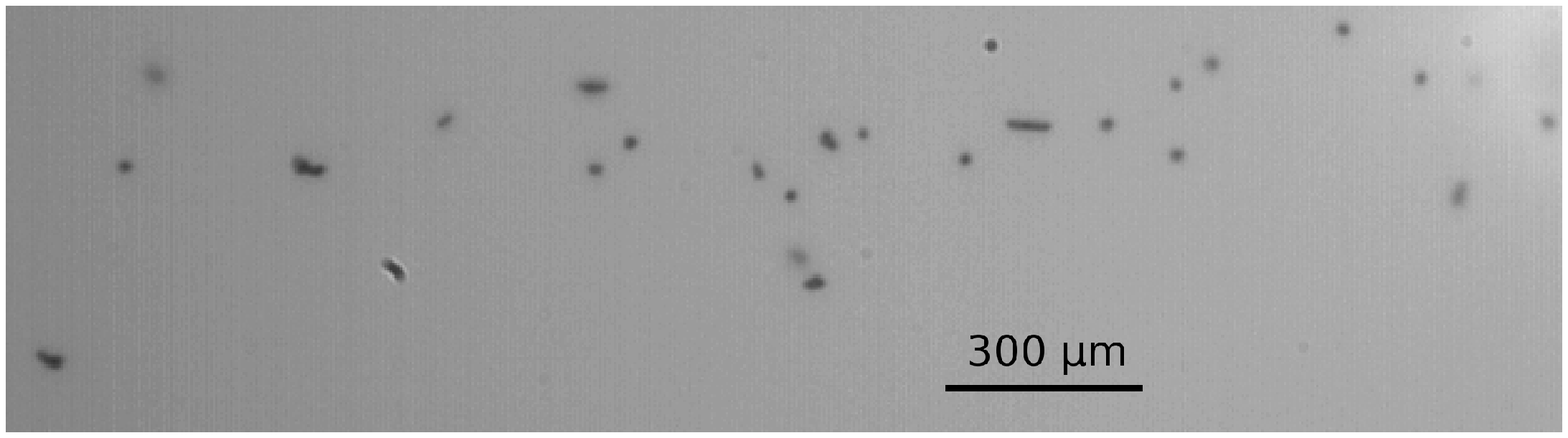}\\
\vspace{2mm}
\includegraphics[width=.48\textwidth, clip=]{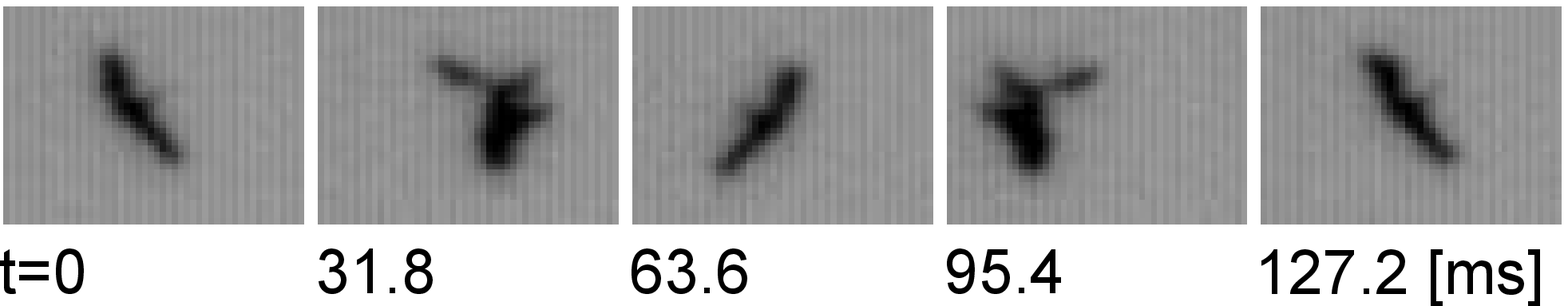}
 \caption{{\bf Upper panel:} Levitating ice particles observed between the Peltier element and the reservoir of liquid nitrogen at a pressure of 1.8\,mbar. About 95\% rotate about their vertical axis. {\bf Lower panel:} Rotation of a single ice aggregate. The image section is $230\,\mu$m$\times 160\,\mu$m.}
 \label{Figlevice}
\end{figure}

To resolve the rotation, we obtain video sequences at 502 frames per second.  Figure~\ref{Figlevice} shows as an example one frame of the recorded video sequences (upper panel) and a rotation sequence of one of the aggregates (lower panel). About 95\% of all the levitating particles are vertically aligned which means that they spin about their vertical axis.  We do not find particles rotating around their horizontal axis. The remaining 5\% of the particles show either no resolvable rotation - they seem to be frozen halfway between the Peltier and the reservoir of liquid nitrogen - or they have an additional velocity component along the \emph{z}-axis, i.e., they are moving up or down and generally show more complex trajectories. We note that in absolute numbers these are only a few particles.

The particles rotating around their vertical axis can be divided into two classes: about half of the rotating particles have an additional component of motion as they move on a horizontal orbit, a rotation bound to the spin of the particles. This is to be expected as the particles are typically of irregular shape (see model below). During a full orbit, they perform one turn around their vertical axis. Furthermore, we see many particles without the additional horizontal movement. However, this might be a resolution effect if the orbit is small.

Our sample contains about 100 ice particles of different sizes and shapes.  Within the resolution limit this includes almost spherical particles as well as cylindric particles, but mostly very irregular-shaped particles. Horizontally extended as well as vertically extended particles -- both rotating around the vertical -- occur (see Fig.~\ref{Figlevice}).
\section[]{Analysis and discussion}
We now discuss two parameters of the ice particles, i.e., their rotation frequency and their shape, in order to estimate their influence on the photophoretic strength.

\subsection{The dependence of photophoresis on the rotation}
In general, particle rotation can change the temperature gradients of illuminated  particles. However, as long as the particles rotate around the vertical (as they do here), always the same side faces the thermal radiation source at the bottom. Such a rotation does \textbf{not} 
change the temperature gradient in a coordinate system fixed to the particle (body fixed). The strength of photophoresis and thermophoresis along the direction of the source of light and the gas temperature gradient do then not depend on the rotation. Therefore, the fact that at least 95\% of all the particles rotate around the vertical is a major finding here. It shows that rotation induced by phoretic forces is not important for the strength of photophoresis and thermophoresis along the vertical. This experimental result clearly shows that systematic rotation does not alter the capability of thermal radiation to lift particles by phoretic forces.

This might no longer be true if an additional sideward illumination and therefore a horizontally directed photophoretic force is added. In protoplanetary discs this might, e.g., be starlight illuminating particles at the surface of a flared disc. The question then remains how the resulting photophoretic strength is influenced by the rotation around the vertical. In case of a strong photophoretic force the rotational state will certainly change, but this is subject to future work. If the vertical force is dominant, we might consider the simple case where rotation would still be tied to the vertical axis. In this case, the rotation frequency would be important. Fast rotation might level out temperature differences and reduce the photophoretic strength induced by the added source of light. On the other hand, slow rotation might lead to a transverse component of the force and accelerate or decelerate a particle, which leads to radial drift.
 
\begin{figure}
\centering
\includegraphics[width=.42\textwidth, clip=]{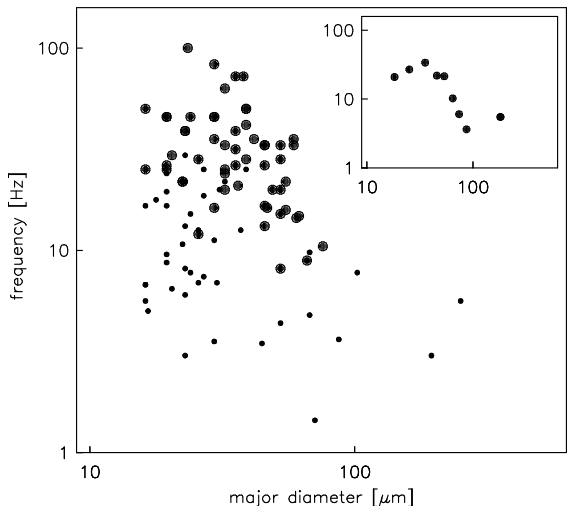}
\caption{{\bf Large panel:} the frequency of rotation of the ice particles is plotted \emph{vs.} their maximum horizontal extent, here called major diameter (both on logarithmic scales). Small symbols represent all particles which show resolvable bound rotation. Large symbols represent all particles where a horizontal orbit does either not exist or is too small to be resolved. The uncertainties of both parameters are small (about the size of the large symbols) and for better visibility not plotted. Small panel: mean values of the frequency of rotation calculated for different ranges of the major diameter (10\,-\,20\,$\mu$m, 20\,-\,30\,$\mu$m, ..., 100\,-\,260\,$\mu$m).}
\label{Figsizevsfreq}
\end{figure}

To quantify such effects, the basic parameter is the rotation frequency. Therefore, we measured this parameter for the levitating particles. While we expect a wide range of frequencies due to the random nature of particle asymmetries, this range might still be confined to and depend on the size of the particle. As a measure of the particle size we chose the horizontal, maximum extent of the ice aggregates because the horizontal extent determines the moment of inertia around the vertical axis. The frequency plotted over the particle size is shown in the large panel of Fig.~\ref{Figsizevsfreq}. Small symbols represent all particles which show resolvable bound rotation. Large symbols represent all particles where a horizontal orbit cannot be resolved. There is no clear trend visible. However, the distributions of particles with and without noticeable bound rotation are offset with respect to their measured frequencies. While particles without bound rotation cover a range of frequencies from roughly 10\,Hz to 100\,Hz, particles with bound rotation need more time for one full turn and spin with a few up to 50\,Hz.

In the small panel of Fig.~\ref{Figsizevsfreq}, we plot the mean of the frequency of rotation \emph{vs.} the corresponding range of diameters (10\,-\,20\,$\mu$m, 20\,-\,30\,$\mu$m, ..., 100\,-\,260\,$\mu$m). The plot suggests that large particles generally tend to have lower frequencies than smaller particles.

We now compare the time a particle needs for a full rotation with the time that is needed for conductive heat transfer in the particle. According to \citet{Krauss2005}, the time-scale is given by $\tau_{\rm heat}=\frac{\rho c a^2}{k_{\rm th}}$, where $\rho$ is the ice density, $c$ is the heat capacity of the ice aggregates, and $k_{\rm th}$ the thermal conductivity. For a typical particle radius of 10\,$\mu$m with a density of 1000\,kg\,m$^{-3}$, a heat capacity of 1000\,J\,kg\,K$^{-1}$, and a thermal conductivity of about 1\,W\,m$^{-1}$\,K$^{-1}$, this characteristic time is 10$^{-4}$\,s or 10000\,Hz. This is sufficiently short to maintain a temperature gradient across the rotating ice particle. An experimental verification of this estimate is subject to future work.
\subsection{The photophoretic alignment of a particle}
As shown by \citet{Kelling2011}, four forces influence the behaviour of the ice aggregates: the thermo- and photophoretic forces which accelerate the particles from the Peltier element towards the reservoir of liquid nitrogen, the gravitational force which opposes the phoretic forces, and gas drag. Gas drag caused by convection within the chamber is not important. However, gas drag is important as friction for particles moving upwards (which are not trapped) and ultimately, gas drag also determines the rotation frequency as it provides damping. In equilibrium, the decelerating torque by gas drag has to balance the accelerating torque by photophoresis and thermophoresis. We note that the difference in temperature inducing phoretic torques and the particle morphology inducing torques by gas drag have not necessarily to be around the same axis. However, within the resolution of our experimental setup, 95\% of all the particles are well described by a vertical rotation axis (see above). If the few particles behaving strangely were induced by such a misalignment, we would expect a more continuous distribution, which we do not observe. One reason might be that damping is proportional to the velocity of a particle surface. Initially, only photophoretic torques act as accelerating force, which dominates the rotation. A non-aligned gas drag torque might therefore not dominate the motion of the particle. This is obviously the case in the experiments, but a theoretical treatment is beyond the scope of this paper.

Thermophoresis and photophoresis are based on the same principle of interaction between gas molecules and differently heated particle surfaces. Therefore, the same physics of particle alignment should work. We then have a situation where gravity acts in one direction and a force depending on a temperature gradient directly opposes gravity. 

\begin{figure}
\centering
\includegraphics[width=.4\textwidth, bb= 0 0 498 821, clip=]{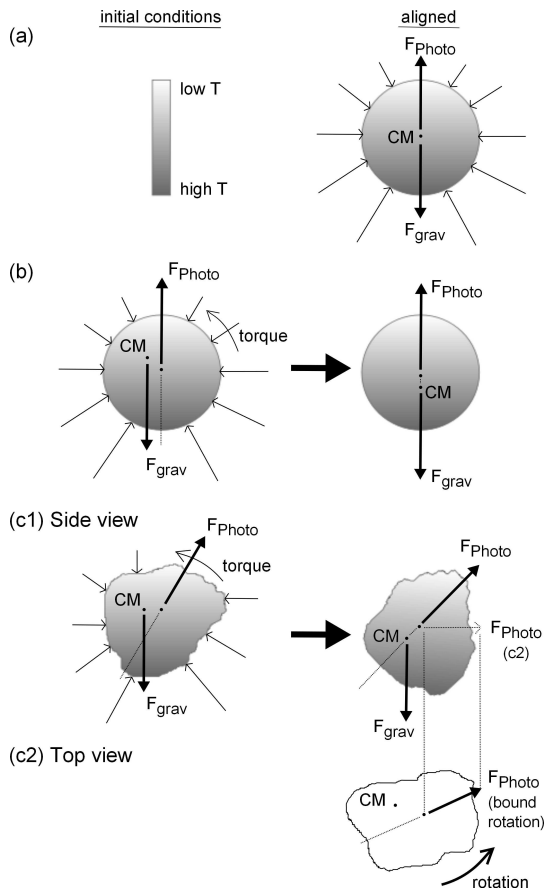}
\caption{Rotation models of ice particles of different morphology. A temperature gradient is induced by a thermal source radiating from below. The centre of mass (CM) is indicated by a black dot, the temperature dependent photophoretic force is marked by slim arrows, the resulting forces by bold arrows. From top to bottom: (a) a homogeneous, spherical particle; (b) a spherical particle where the centre of mass is offset (left); this induces torque which causes rotation until the centre of the particle and the centre of mass are aligned (right); (c) a spherical particle with surface structure seen from the side (c1) and from above (c2); again, a state of equilibrium can be established (right).}
\label{Figrotmodel}
\end{figure}

Rotation due to the different torques and forces is a problem which is highly complex and cannot easily be treated theoretically. Nevertheless, we here provide a simple view why particles should align to and rotate around the vertical (see Fig.~\ref{Figrotmodel}): in terms of basic physics, the motion of a particle can be divided into the motion of the centre of mass due to the total force and a rotation around the centre of mass due to a torque around the centre of mass. To understand particle rotation, we change its shape from a perfect sphere to a realistic, asymmetric particle in three steps. 

Step 1: often considered, but rarely realised by nature are homogeneous, spherical particles. Due to its symmetry, a homogeneous, spherical particle does not experience any torque and will not rotate (or at least, any initial rotation will quickly be damped by gas friction). Phoretic forces are directed upwards and for a levitating particle the phoretic forces balance gravity (Fig.~\ref{Figrotmodel}~a). 

Step 2: as a next step on the way to a more realistic shape we consider a spherical particle where the centre of mass is not the geometrical centre of the sphere. Such a particle will also experience a phoretic force upwards which might be balancing gravity. However, in addition the particle is subject to a torque (Fig.~\ref{Figrotmodel}~b, left). This particle will rotate until the centre of the particle and the centre of mass are aligned to the vertical. In that configuration, the torque vanishes. The particle might oscillate a few times but due to gas damping, the oscillation will subside and the particle will be aligned eventually (Fig.~\ref{Figrotmodel}~b, right). 

Step 3: adding surface structure to the particle leads to additional torques due to the fact that the phoretic forces are normal to the surface elements. Therefore, photophoresis does no longer only act vertically, but the net effect also includes a horizontal component (Fig.~\ref{Figrotmodel}~c1, left). Nevertheless, also in this case a new configuration will be established where torques around the vertical vanish (Fig.~\ref{Figrotmodel}~c1, right). However, a torque around a horizontal axis can prevail. If gravity is balanced by the vertical component of the photophoretic force, a torque around the vertical due to irregularities of the surface will usually be left, for which we also indicate a top view here (Fig.~\ref{Figrotmodel}~c2). 

As mentioned before, rotation around the vertical does not change the top and bottom parts of the particle. With respect to a coordinate system fixed to the particle, this torque remains constant as illustrated in Fig.~\ref{Figrotmodel}~c and will accelerate the particle around the vertical axis. 

So far, we concentrated on the torque, but in addition to torques we also have to consider that non-sphericity induces a total force acting horizontally.  
If tuned in a way that the vertical component is balanced by gravity, only the sideward motion remains. 
If the particle is vertically aligned, this force becomes body fixed, i.e., is
tied to the particle coordinate system. If the particle rotates, so will the direction of the horizontal force. This naturally leads to a bound rotation where the orbit of the centre of mass has the same frequency as the spin of the particle.

Our model ignores any precession which might result from differences between angular momentum and rotation velocity and which might misalign the axis of the particle. Rotational damping by gas drag might also misalign the particle as gas drag is not necessarily inducing a torque around the same axis. However, as long as the differences are small, these misalignment effects will be superimposed and dominated by the main phoretic torque. 

Slowly rotating bodies imply that the bound rotation is also slow and particle
orbits can be larger. This is consistent with our observations that bound rotation (motion of the centre of mass) takes place preferentially for slow rotators. The fact that there is no visible sign of misalignment for bound rotators, indicates that gas drag by horizontal motion does at least to a certain extent not influence the alignment. 
\section{Conclusions}

 Our experiments show how particles rotate under the influence of a photophoretic force, for which our model gives a plausible explanation. As the rotation axis aligns to the direction of illumination, the rotation, independent of its frequency, does not change the strength of photophoresis along the line of radiation. At the given (slow) rotation frequencies, temperature gradients appear to adjust rapidly enough to allow additional photophoretic forces from secondary radiation sources. However, a verification of this theory is subject to future work.

In general, our work supports the concept of photophoretic transport of levitating particles over the surface of protoplanetary discs \citep[][]{Wurm2009b,Wurm2009a}.

\section*{Acknowledgments}

The authors would like to thank the anonymous referee for helpful comments which improved the quality of this paper. Additionally, we wish to thank Stefan Hagenacker for his support with the laboratory work.

\bibliographystyle{mn2e}
\bibliography{bibliography}

\begin{thebibliography}{}

\bibitem[\protect\citeauthoryear{Kelling, Wurm \& D\"urmann}{Kelling
  et~al.}{2011}]{Kelling2011}
Kelling T.,  Wurm G.,    D\"urmann C.,  2011, {Tracing ice particles with
  temperature gradients at mbar pressure}, Review of Scientific Instruments, in
  revision

\bibitem[\protect\citeauthoryear{{Krauss} \& {Wurm}}{{Krauss} \&
  {Wurm}}{2005}]{Krauss2005}
{Krauss} O.,  {Wurm} G.,  2005, ApJ, 630, 1088

\bibitem[\protect\citeauthoryear{{Krauss}, {Wurm}, {Mousis}, {Petit}, {Horner}
  \& {Alibert}}{{Krauss} et~al.}{2007}]{Krauss2007}
{Krauss} O.,  {Wurm} G.,  {Mousis} O.,  {Petit} J.-M.,  {Horner} J.,
  {Alibert} Y.,  2007, A\&A, 462, 977

\bibitem[\protect\citeauthoryear{{Wurm} \& {Haack}}{{Wurm} \&
  {Haack}}{2009a}]{Wurm2009b}
{Wurm} G.,  {Haack} H.,  2009a, in {T.~Henning, E.~Gr{\"u}n, \& J.~Steinacker}
  ed., Cosmic Dust - Near and Far Vol.~414 of Astronomical Society of the
  Pacific Conference Series, {Levitation of Dust at the Surface of
  Protoplanetary Disks}.
pp 509--+

\bibitem[\protect\citeauthoryear{{Wurm} \& {Haack}}{{Wurm} \&
  {Haack}}{2009b}]{Wurm2009a}
{Wurm} G.,  {Haack} H.,  2009b, Meteoritics and Planetary Science, 44, 689

\end{thebibliography}

\label{lastpage}

\end{document}